\documentstyle[12pt]{article}

\newcommand{\beq}{\begin{equation}}
\newcommand{\eeq}{\end{equation}}
\newcommand{\fie}{\varphi}

\newcommand{\tet}{\vartheta}

\newcommand{\al}{\alpha}

\newcommand{\de}{\delta}
\newcommand{\De}{\Delta}
\newcommand{\la}{\lambda}
\newcommand{\La}{\Lambda}
\newcommand{\om}{\omega}

\newcommand{\ep}{\epsilon}
\newcommand{\ga}{\gamma}

\newcommand{\si}{\sigma}

\newcommand{\ti}{\tilde}
\newcommand{\Up}{\Upsilon}
\newcommand{\up}{\upsilon}

\newcommand{\bdm}{\begin{displaymath}}
\newcommand{\edm}{\end{displaymath}}

\begin{document}

\title{Nonlinear stability analysis of plane Poiseuille flow by normal forms}

\author{A. Rauh,  T. Zachrau, and J. Zoller \\
     Fachbereich Physik, Carl von Ossietzky Universit\"at Oldenburg,\\
        D-26111 Oldenburg, Germany}

\maketitle

\begin{abstract}
In the subcritical interval of the Reynolds number
$4320\leq R\leq R_c\equiv 5772$,
the Navier--Stokes equations of the two--dimensional plane Poiseuille flow
are approximated by a 22--dimensional Galerkin representation formed from
eigenfunctions
of the Orr--Sommerfeld equation. The resulting dynamical system is brought
into a
generalized normal form which is characterized by a disposable parameter
controlling the magnitude of denominators of the normal form transformation.
As rigorously proved, the generalized normal form decouples into a
low--dimensional
dominant and a slaved subsystem. From the dominant system the critical
amplitude is
calculated as a function of the Reynolds number. As compared with the Landau
method,
which works down to $R=5300$,
the phase velocity of the critical mode agrees within $1\%$; the
critical amplitude is reproduced similarly well except close to the critical
point, where the maximal error is about $16\%$.
We also examine boundary conditions which partly differ from the usual ones.
\end{abstract}

\section{Introduction}

In local bifurcation theory the center manifold \cite{Carr},
Landau \cite{Landau,Stuart}, and normal form
\cite{Arnold,GuckenheimerH,Wiggins} methods have been invaluable tools.
As a common
feature, a dynamical system is approximated by series expansions around
the fixed point in phase space, and in the case of the first two methods also
around the critical point in parameter space. In many cases
rigorous theorems are available to ensure the correct behaviour of the system
in a sufficiently small neighbourhood of the critical point.

In this article we present a generalized normal form scheme with the aim
to explore the subcritical , nonlinear regime in a finite neighbourhood
of the fixed point.
The method is characterized by a disposable parameter $\ep$ which controls
the magnitude of the denominators of the normal form transformation.
The case $\ep=0$
corresponds to the most usual normal form where resonance denominators are
avoided.
With $\ep>0$ we additionally  discard quasi--resonant denominators with
absolute magnitude $\leq\ep$.
While increasing $\ep$ enlarges the definition domain of the transformation,
the normal form
gets more and more complicated by the occurrence of further nonlinear terms.
Nevertheless, as rigorously proved under rather general conditions on $\ep$,
the normal form decouples into a
low--dimensional dominant and a slaved \cite{Haken} subsystem in the sense
that stability of the dominant system
induces stability of the remaining degrees of freedom. The theorem proved
generalizes previous work \cite{RauhP,PadeRT}.

We apply the method to the well studied subcritical stability analysis of the
plane Poiseuille
flow, see  \cite{Stuart,Watson,HerbertL,Ven,Herbert,OrszagP,Bayly,Eckhaus}.
The Navier--Stokes equations are approximated by a Galerkin representation in
the basis
of 22 eigenfunctions of the linearized part, i.e. the Orr--Sommerfeld equation.
The Galerkin system, with diagonal linear part, is then subject to a normal
form transformation  with  the resonance condition generalized through the
interval parameter $\ep$. The latter is chosen so as to obtain a
three--dimensional dominant system
for the normal form in the Reynolds number interval $4320\leq R\leq R_c$.
The dominant system exhibits a Hopf--bifurcation at a critical amplitude
$r_c\equiv r_c(R)$.
The normal form series expansion is carried out up to cubic monomials.

As compared with previous investigations of the plane Poiseuille flow
\cite{HerbertL,OrszagP,Eckhaus}
we examine the effect of slightly different boundary conditions: for the mode
with zero wave number $q$
in downstream direction we adopt the same boundary conditions as for the other
modes.
Because the different modes are mixed by the nonlinear interaction, we think
one should stay in the same function space for consistency.

To establish the Orr--Sommerfeld eigenfunctions of the linear problem
we adopt the usual scheme of Chebyshev polynomial expansions \cite{Orszag}.
We use 50 even and 50 odd polynomials.
Eventually the critical curve $r_c(R)$ obtained from the normal form is
transformed back into
the original phase space and compared with the  Landau method \cite{Ven}.

The definition domain of the method is principally determined
by the condition that the Jacobian $J$ of the normal form transformation
differs from zero
in an open neighbourhood of the fixed point, where $J=1$. Because of our
cut--off at cubic terms
the corresponding $J$ does not allow for a reliable extrapolation.
In the given case we check the definition domain in a pragmatic way
by integrating the Galerkin
system directly for starting points around the critical curve found by the
normal form method.

We set $\al=\al_c=1.02$ always, which is the critical value of the basic wave
number in the downstream direction at the critical Reynolds number $R=R_c$.

The article is organized as follows: In the following section the
two--dimensional Navier--Stokes
equations are expressed  in terms of the scalar stream function $\Psi(x,z)$.
The numerical methods to obtain the eigenfunctions of the linearized
stationary part are briefly sketched and the new
boundary conditions are stated.
In section 3 the Galerkin representation is derived.
In the next section our generalized normal form scheme is introduced and a
decoupling
 theorem is formulated which is proved in Appendix B. The fifth section
contains the application
 of the normal form scheme to the nonlinear stability analysis of the plane
Poiseuille flow.
In particular, the detailed normal form with numerical coefficients is given
at the lowest
Reynolds number considered $R=4320$. In section 6 the quantitative results are
 presented and
compared with the results of the Landau method \cite{Ven}. This is followed by
 a conclusion section.

\section{Basic scheme}
The two--dimensional velocity field~$(u,w)$ of the
incompressible plane Poiseuille flow is most conveniently
described in terms of the stream function $\Psi$ as
\beq
\label{stf}
u=\frac{\partial\Psi}{\partial z}\;\;\;\; ;
                \;\;\;\; w=-\frac{\partial\Psi}{\partial x}
\eeq
where $u$ denotes the streamwise velocity component and $w$ the
component normal to the boundaries. The Navier--Stokes equation
then reads \cite{LandauL,Schlichting}
\beq
\label{a1}
\frac{\partial}{\partial t}\nabla^2\Psi +
\frac{\partial \Psi}{\partial z}
\frac{\partial \nabla^2\Psi}{\partial x}
- \frac{\partial \Psi}{\partial x}
\frac{\partial \nabla^2\Psi}{\partial z}
=\frac{1}{R}\nabla^4\Psi\;.
\eeq
The stream function~$\Psi$ is decomposed as \( \Psi=\Psi^b +\Psi^d \), where
the basic flow  $\Psi^b=z-(1/3)z^3$ and the disturbance field~$\Psi^d$ is
written as a superposition of plane waves
\beq
\label{ans}
\Psi^d(x,z,t)=\sum_{q=-\infty}^{\infty}\Psi_q(z,t) \exp(iq\al x)
\eeq
with basic wave number $\al$. All magnitudes are dimensionless:
the coordinates are measured in terms of the channel half--width $h$,
$(u,w)$ by the maximal unperturbed velocity $U_0$. The Reynolds number
$R=U_0h/\nu$ where $\nu$ denotes the kinematic viscosity. With the
ansatz (\ref{ans}), equation (\ref{a1}) takes
the following well--known form \cite{Stuart}:
\bdm
L\Psi_{q}(z,t)-R\frac{\partial}
            {\partial t}(D^2-q^2\alpha^2)\Psi_q(z,t)=
\edm
\bdm
- iR\al\sum_{q'=-\infty}^\infty
[(q-q')\Psi_{q-q'}(z,t)(D^3-q'^2\al^2D)\Psi_{q'}(z,t)
\edm
\beq
\label{basnl}
-q'D\Psi_{q-q'}(z,t)(D^2-q'^2\al^2)\Psi_{q'}(z,t)]\;,
\eeq
where
\beq
\label{L}
L=(D^2-q^2\alpha^2)^2-i\al qR[2+U(z)(D^2-q^2\al^2)]\;,
\eeq
with $D=\partial/\partial z$ and $U(z)=1-z^2$.

The linearized, stationary part of (\ref{basnl}) with
$\Psi_q(z,t)=\Phi_q(z)\exp(\la_q t)$ leads to the Orr--Sommerfeld equation
\beq
\label{lep}
L\Phi_q(z)=\la_q R(D^2-q^2\alpha^2)\Phi_q(z)\;.
\eeq

We adopt the boundary conditions
\beq
\label{bcnew}
\Phi_q (z=\pm 1)=D\Phi_q (z=\pm 1)=0
\eeq
for all $q=0,\pm 1,\ldots$. In the case $q=0$ this is at variance
with other authors \cite{Herbert,OrszagP,Eckhaus} who prefer
\beq
\label{bcold}
D\Phi_{q=0} (z=\pm 1)=D^2\Phi_{q=0} (z=\pm 1)=0\;.
\eeq
We think the latter condition is inconsistent in the nonlinear case where
it would lead to a mixing of functions with different boundary conditions.
As is discussed, e.g. in \cite{SoibelmanM} the boundary conditions
(\ref{bcnew})
implies that the mass flux is kept constant whereas the case
(\ref{bcold}) refers to
an experiment with constant pressure gradient averaged over the channel width.

For the normal form transformation to follow we need the linearized
part of (\ref{basnl}) in diagonal form, if possible. Therefore we
expand the functions $\Psi_q$ in terms of the eigenfunctions $\Phi_{q,\nu}$
of the Orr--Sommerfeld equation, henceforth abbreviated by OSF
\beq
\label{ansnl}
\Psi^d(x,z,t)=
\sum_{q,\nu}\;\eta_{q,\nu}(t)\Phi_{q,\nu}(z)\exp (iq\al x)\;,
\eeq
where $\nu=1,2,\ldots$ numbers the different eigenfunctions for given
wave number $q=0,\pm 1,\pm 2,\pm 3,\ldots$.

We solve the Orr--Sommerfeld equation numerically with standard procedures
proposed in \cite{Orszag,GaryH} using Chebyshev polynomials as basis functions.
We took $50$ polynomials both for the symmetric and the antisymmetric OSF.
As a check of our $IMSL$ procedure we reproduced the eigenvalues given
in \cite{Orszag} for the case $\al=1$, $R=10000$.

The case $q=0$ can be solved analytically. Because of our unified
boundary conditions (\ref{bcnew}) we now get only antisymmetric functions,
namely
\beq
\label{eig0}
\Phi_{0,\nu}(z)=B_\nu[\sin(k_\nu z)-z\sin(k_\nu)]
\;;\;\;\la_{0,\nu}=-k_\nu^2/R
\;;\;\;k_\nu=\tan(k_\nu)\;.
\eeq
In the usual case (\ref{bcold}) both symmetric and antisymmetric functions
show up.
However, one applies a kind of superselection rule suggested by numerical
experience
\beq
\label{selrul}
\Phi_{q,\nu}(z)=(-1)^{q+1}\Phi_{q,\nu}(-z)\;,
\eeq
which excludes symmetric functions in the case $q=0$.
Our antisymmetric functions differ from the traditional ones
\beq
\label{q0old}
\tilde{\Phi}_{0,\nu}(z)=C_{\nu}[\sin(\nu\pi z)+\nu\pi z(-1)^{\nu+1}]
\;\;;\;\;\la_{0,\nu}=-\nu^2\pi^2/R\;.
\eeq
Clearly, this does not affect the linear stability analysis because the
critical amplitude is in the space with $q=\pm 1$. On the other hand,
the different functions $\Phi_{0,\nu}$ should have an effect upon the
nonlinear stability properties.

Later on we measure the strength of the disturbance field in terms of the
mean energy
\bdm
E(\eta)=\frac{1}{2}\overline{ u^2+w^2}:=\frac{\al}{8\pi}
       \int_{-1}^1dz\int_0^{\frac{2\pi}{\al}}dx(u^2+w^2)=
\edm
\beq
\label{ave}
    =\frac{1}{2}\sum_{q,\nu,q',\nu'}\eta_{q,\nu}(t)\eta_{q',\nu'}(t)
         M_{q\nu,q'\nu'}\;,
\eeq
which involves the metric components of the eigenfunctions as
\beq
\label{M}
M_{q\nu,q'\nu'}=
\frac{1}{2}\left[-qq'\al^2
         \int_{-1}^1dz\Phi_{q,\nu}\Phi_{q',\nu'}
         +\int_{-1}^1\;dz
\frac{d\Phi_{q,\nu}}{dz}\frac{d\Phi_{q',\nu'}}{dz}\right]
\de_{q',-q}\;,
\eeq
where $\de$ denotes the Kronecker symbol.
\section{ Galerkin representation}
In order to derive the Galerkin system for (\ref{basnl}) we need
both the right and left eigenfunctions of the Orr--Sommerfeld
equation. It is convenient to introduce the abbreviations
$j=(q,\nu)$, $k=(q',\nu')$ and $l=(q'',\nu'')$, and the index projection
$[j]=q$, $[k]=q'$ and $[l]=q''$ and furthermore
\beq
\ti{\Phi}_k(z)=D\Phi_k(z)\;;\;
G_k(z)=(D^2-[k]^2\al^2)\Phi_k(z)\;;\;
\ti{G}_k(z)=DG_k(z)\;.
\eeq
The Chebyshev expansion coefficients of
$\Phi_k(z)$, $\ti{\Phi}_k(z)$, $G_k(z)$, $\ti{G}_k(z)$ are denoted by
$a_n^k$, $\ti{a}_n^k$, $g_n^k$, $\ti{g}_n^k$, respectively.
By comparing the coefficients of the independent functions $\exp (iq\al x)$,
we obtain
\beq
\label{basnl2}
\sum_\nu[\frac{d\eta_j}{dt}-\la_j\eta_j]
(D^2-q^2\al^2)\Phi_j(z)=
i\al\sum_{k,l}\eta_k\eta_l[
\Up_{kl}(z)-\ti{\Up}_{kl}(z)]\de_{[k]+[l],[j]}
\eeq
where $\Up_{kl}(z)=[k]\Phi_k(z)\ti{G}_l(z)$ and
$\ti{\Up}_{kl}(z)=[l]\ti{\Phi}_k(z)G_l(z)$.
 The corresponding coefficients
$\up_n^{kl}$ and $\ti{\up}_n^{kl}$ are found according to the product rule
given in \cite{Orszag}
\beq
\label{up}
\up_n^{kl}=\frac{[k]}{2c_n}\sum_{n'=-\infty}^\infty\;
\bar{a}_{n-n'}^k\bar{\ti{g}}_{n'}^l
\;;\;
\ti{\up}_n^{kl}=\frac{[l]}{2c_n}\sum_{n'=-\infty}^\infty\;
\bar{\ti{a}}_{n-n'}^k\bar{g}_{n'}^l
\;;\;\;\;n=0,1,\ldots
\eeq
with $\bar{a}_n=c_{|n|}a_{|n|}$ and $c_n=1+\de_{n,0}$. Furthermore, the
operator $(D^2-q^2\al^2)$ is represented by the square matrix ${\cal D}^q$ as
\beq
(D^2-q^2\al^2)T_n(z)
=\sum_{n'}\;{\cal D}_{n'n}^qT_{n'}(z)\;,
\eeq
where $T_n(z)$ denotes the Chebyshev polynomials of order $n=0,1,\ldots$.
If $q\neq 0$, then ${\cal D}^q$ is nonsingular. Its elements can be inferred
from formula (A5) given in \cite{Orszag}.

Comparing the coefficients of $T_n(z)$ in (\ref{basnl2}) we can write
\beq
\label{basnl3}
\sum_\nu[\frac{d\eta_j}{dt}-\la_j\eta_j]
\sum_{n'}{\cal D}^q_{nn'}a^j_{n'}=
i\al\sum_{k,l}\eta_k\eta_l[
\up^{kl}_n-\ti{\up}^{kl}_n]\de_{[k]+[l],[j]}\;.
\eeq
In the case $q\neq 0$ we multiply (\ref{basnl3}) from the left by
$({\cal D}^q)^{-1}$ and then by the 1--row matrix $b^j$ formed by the
Chebyshev expansion coefficients $b^j_n$ $n=0,1,\ldots$ of the left
eigenfunction $\Phi^L_j(z)$.
This gives rise to the nonlinear Galerkin system we are looking for
\beq
\label{gals}
\frac{d}{dt}\eta_j=\la_j\eta_j+\sum_{k,l}
\;N_{j|kl}\eta_k\eta_l\;,
\eeq
with the constant coefficients
\bdm
N_{j|kl}=i\al\frac{\sum_n\;b_n^j
         \sum_{n'}({\cal D}^q)^{-1}_{nn'}
               [\up_{n'}^{kl}-\ti{\up}_{n'}^{kl}]}
         {\sum_n\;b_n^ja_n^j}\de_{[k]+[l],[j]}\;;
\edm
\beq
\label{njkl}
 q\neq 0\;\;;\;\;j=(q,\nu)\;\;;\;\;
 k=(q',\nu')\;\;;\;\;l=(q'',\nu'')\;.
\eeq
The denominator is obtained from the left
and right eigenfunctions as delivered by the $IMSL$
routine.

In the case $q=0$ the left and the right eigenfunctions $\Phi_{0,\nu}(z)$,
given through (\ref{eig0}), are the same. We denote the corresponding
expansion coefficients by $d_n^j$ with $j=(0,\nu)$. The $N_{j|kl}$ are now
given by
\beq
\label{njkl0}
N_{j|kl}=
i\al\frac{\sum_n\;d_n^j(\up_n^{kl}-\ti{\up}_n^{kl})}
         {\sum_n\;d_n^jd_n^j}\de_{[k]+[l],[j]}
\;\;\;;\;\;\;j=(0,\nu)\;.
\eeq
Of course, for all $j$ only the symmetric part $(N_{j|kl}+N_{j|lk})/2$ enters
(\ref{gals}). By the reality of the Navier--Stokes equations
we have the property
\beq
N_{\bar{j}|\bar{k}\bar{l}}=N^\ast_{j|kl}\;,
\eeq
where the star denotes complex conjugation and  $\overline{(q,\nu)}=(-q,\nu)$.
As an implication the amplitudes $\eta_j$ obey the relation
\beq
\label{etacc}
\eta_{\bar{j}}=\eta^\ast_j\;.
\eeq
By the Galerkin form (\ref{gals}) we tacitly assumed that the linear problem
can be fully diagonalized. A degeneracy can be avoided by
slightly shifting the Reynolds number $R$.

\section{ Interval normal form}

By means of a normal form transformation $\eta\rightarrow\xi$
we now decouple the Galerkin system (\ref{gals}) into a low--dimensiomal
dominant system and a slaved subspace.
The nonlinear, near identity transformation is written in multi--index
notation as
\beq
\label{trans}
\eta_j=\xi_j+\sum_{m\in M} A_j(m)\xi^m\;\;;
                       \;\;A_j(m)\in {\bf C}\;\;;
                       \;\;j=1,2,\ldots ,n\;,
\eeq
where $\xi^m=\xi_1^{m_1}\xi_2^{m_2}\cdots \xi_n^{m_n}$ is a monomial of order
$|m|=m_1+m_2+\ldots+m_n$ and $M$ is the set of non--negative integer vectors
\beq
\label{m}
M:=\{ m|\; m\in {\bf N}_0^n\;\;;\;\; |m|\geq 2\}\;.
\eeq
The transformed version of the Galerkin system then has the general form
\beq
\label{dotxi}
\dot{\xi}_j=\la_j\xi_j+\sum_{m\in M}\;B_j(m)\xi^m\;,
\eeq
which we try to simplify in the spirit of the normal form transformation
as far as possible.
In practical situations like our present case it is important to avoid not
only zero denominators but also near resonances, otherwise the definition
domain of the transformation (\ref{trans}) could become unacceptably small.
In view of this we introduce an interval parameter $\ep$ and define
the generalized resonant set ${\cal R}_j(\ep)$. It determines the set of
monomials to be kept in (\ref{dotxi}) as follows
\beq
\label{calR}
{\cal R}_j(\ep):=\{m\;|\;d_j(m)\leq\ep,\;m\in M\}\;.
\eeq
Here $d_j(m)$ is the absolute magnitude of a denominator of the normal
form scheme
\beq
\label{Djm}
D_j(m):=(m,\la)-\la_j\;;\;
(m,\la)=m_1\la_1+m_2\la_2+\ldots+m_n\la_n
\;;\;\;\;m\in M\;.
\eeq
The coefficients
$A_j(m)$ and $B_j(m)$ of (\ref{trans}) and of the
normal form (\ref{dotxi}), respectively,
are connected by recurrence relations of the following structure
\beq
\label{Ajm}
A_j(m)=-\frac{1}{D_j(m)}  [ B_j(m)+ P( A_k(m'), B_l(m'')) ] \,,
\eeq
where $P$ is a polynomial which contains only coefficients $A_k(m')$ and
$B_l(m'')$ of lower order $|m'|,|m''|<|m|$.
If $m\in{\cal R}_j(\ep)$ then $d_j(m)\equiv |D_j(m)|\leq\ep$ and
one sets the square bracket in (\ref{Ajm}) equal to zero with
$ B_j(m)=-P( A_k(m'), B_l(m''))$. Otherwise $B_j(m)$ is set equal to zero.
In Appendix A we state the
explicit relations for $|m|=2$ and $|m|=3$; the general recurrence
relations can be found  in \cite{RauhP}.

In qualitative normal form studies, see e.g.
\cite{Arnold,GuckenheimerH,Wiggins,ArneodoT} the
interval parameter $\ep$ is set equal to zero. With $\ep\neq 0$ we now
get the more complicated normal form
\beq
\label{dotxi2}
\dot{\xi}_j=\la_j\xi_j+\sum_{m\in {\cal R}_j(\ep)}\;B_j(m)\xi^m\;,
\eeq
with more nonlinear monomials as compared with the pure case $\ep=0$.

Nevertheless, with a suitable choice of $\ep$ the interval normal form
(\ref{dotxi2}) exhibits a useful decoupling property.
To reveal this, we partition the set of eigenvalues $\la_j$ into
groups characterized by their real parts.
To this end the left hand complex
plane of eigenvalues  is divided into stripes which are parallel to the
imaginary axis and of different width according to some convenient choice.
The first stripe
contains the least damped eigenvalues, i.e. with smallest modulus of the
real part of $\la_j$.
The second stripe is empty. The third stripe
contains the second eigenvalue group. The fourth interval is empty, and so on.
For illustration see fig. 1.

\vspace{1cm}

{\bf Here insert fig. 1}

\vspace{1cm}

We label the groups by the index
$\si$ and  the members of one group by the index $\tau$.
As should be noted, we have now three different equivalent labellings for the
OSF:
i) $j=(q,\nu)$ where $q=0,\pm 1,\pm 2, \ldots$ refers to the different
streamwise wave numbers and $\nu=1,2,\ldots$ numbers the different states at
a given $q$;
ii) $j=1,2,\ldots$ which numbers the different OSF ordered e.g. by
decreasing real parts $\Re (\la_j)$;
iii) $j=(\si\tau)$ which is the labelling
regarding decoupling, specifically
\beq
\label{lmn}
\la_j=\rho_{\si\tau}+i\om_{\si\tau}\;;\;\;
\si=1,\ldots,\hat{n}\;;\;
\tau=1,\ldots,E_\si\;;\;
j=1,\ldots ,\sum_{\si=1}^{\hat{n}}E_\si=n\;.
\eeq
Note that the real parts $\rho_{\si\tau}<0$ in this study.
By convention, $\tau=1$ and $\tau=E_\si$ denote the maximum and minimum of
the real parts for a given group $\si$, respectively. The same labelling is
carried over to $m_j\rightarrow m_{\si\tau}$ and
$\xi_j\rightarrow\xi_{\si\tau}$.
Furthermore, we define the subspaces $\Xi_\si$ which correspond to the
eigenvalues contained in one group:
\beq
\Xi_\si:=(\xi_{\si 1},\xi_{\si 2},\ldots,\xi_{\si E_\si})\;.
\eeq

We now define the width $W_\si$ of a group by
\bdm
W_1=|\rho_{1E_1}|
\edm
\beq
\label{width}
W_\si= |\rho_{\si E_\si}|-|\rho_{\si 1}|\;\;\;{\rm for}\;\;\si\geq 2
\eeq
and the distance $\De_\si$ between consecutive stripes containing
eigenvalues by
\beq
\label{dist}
\De_\si=|\rho_{\si+1,1}|-|\rho_{\si E_\si}|\;.
\eeq
Furthermore  $\La_\si$ denotes an $E_\si\times E_\si$ matrix and
$P_\si$ a polynomial which starts with quadratic terms or is identically
zero.
As it turns out, the interval normal form decouples substantially provided
the distances $\De_{\si}$ between different groups are sufficiently large
and the width $W_\si$ of a group is sufficiently small. The precise
theorem reads

{\bf Decoupling theorem}:
If the interval parameter $\ep$ allows for the properties
\beq
\label{cond1}
\De_{\si}>\ep-|\rho_{11}|\;\;\;
{\rm for}\;\;\si=1,2,\ldots\;,
\eeq
and
\beq
\label{cond2}
W_\si<|\rho_{\si 1}|-\ep\;\;\;
{\rm for}\;\;\si=2,3,\ldots\;,
\eeq
then the normal form {\rm (\ref{dotxi2})} has the following structure
\beq
\label{dynsys1}
\dot{\Xi}_1 =  \La_1\Xi_1       + P_1(\Xi_1)  \;;
\eeq
\beq
\label{dynsys2}
\dot{\Xi}_\si  = \La_\si(\Xi_1,\ldots,\Xi_{\si-1})\Xi_\si +
           P_\si(\Xi_1,\ldots,\Xi_{\si-1})\;;\;\;\;
              \si=2,\ldots,\hat{n}\;.
\eeq

The proof of this theorem is given in Appendix~B. As should be remarked, in
practical application it is advantageous to consider also the imaginary parts
of the eigenvalues, which leads to a smaller resonance set ${\cal R}_j(\ep)$
and thus simplifies the normal form, in general.

The structure of (\ref{dynsys1}), (\ref{dynsys2}) implies that the
equations of the higher subspaces are subsequently linear with time dependent
coefficients, if the solutions of the preceding subspaces are known.
Moreover, if the subsystem $\Xi_1$ has a stable fixed point at zero,
then the fixed point is globally
attractive for arbitrary initial values  of the higher subspaces
$\Xi_\si(t=0)$ with $\si=2,\ldots,\hat{n}$ provided the initial values of the
subspace $\Xi_1$ are in the basin of attraction of its fixed point. This
is exemplarily proved in Appendix C by means of a Lyapunov--type function.
The theorem is formal as long as  the existence domain of the transformation
(\ref{trans}) is not established. If all eigenvalues are damped, then a finite
convergence radius of (\ref{trans}) and thus a finite domain with Jacobian
$J\neq 0$ can be expected; in \cite{RauhP} this was proved for a special
choice of $\ep\neq 0$.

In the subcritical region with negative real parts, the condition
(\ref{cond1}) can always be fulfilled, provided the interval parameter $\ep$
is chosen small enough as $0<\ep\leq |\rho_{11}|$.
In this case (\ref{cond2}) is unnecessary, because there is no
condition on the mutual distances between different groups so that we
can always choose $W_\si=0$. This situation was considered in \cite{RauhP}
where the system $\Xi_1$ is shown to contain only quadratic nonlinearities.
However, with such a choice the parameter $\ep$ would become
arbitrarily small in the limit $R\rightarrow R_c$, with the implication that
the definition domain of (\ref{trans}) may shrink to zero.

In our application we choose  $\ep>|\rho_{11}|$.
The problem is now finding a suitable partitioning into groups which is
compatible with the conditions (\ref{cond1}), (\ref{cond2}).
Moreover, it is desirable to maintain a feasible form of the dominant
subsystem $\Xi_1$. With the given finite cut--off of
22 eigenfunctions we have no problems finding a convenient partitioning.

\section{ Application to Poiseuille flow}

We choose the interval parameter $\ep$ as
\beq
\label{ep}
\ep=\max\{|\la_{q=0,\nu=1}|,2|\Re (\la_{q=1,\nu=1})|\}
\eeq
\bdm
\ep<|\Im (\la_{q=1,\nu=1})|\;,
\edm
where $\Re$ and $\Im$ denote the real and imaginary part, respectively.
This choice provides us with a three--dimensional dominant subsystem
in the Reynolds number interval $4320\leq R\leq R_c$. The
partitioning of eigenvalues into groups and with it the structure of the
normal form is the same in this $R$-interval.
The situation is quantitatively shown in fig. 1 for $R=4320$.
It is compatible with the conditions (\ref{cond1}) and (\ref{cond2}) for
the distances $\De_\si$ and widths $W_\si$ of the stripes.
As should be noticed, our chosen
interval parameter $\ep$ increases with decreasing Reynolds number $R$
according to the $R$--dependence of the eigenvalues.

The first stripe contains one real and the two complex conjugate eigenvalues
which become critical at $R=R_c$.
For convenience we denote the three dominant variables and eigenvalues
by $(\xi_1,\xi_2,\xi_3)$ and $(\la_1,\la_2,\la_3)$, respectively.

Because $\la_1=\la_2^\ast$ and $\la_3=\la_3^\ast$ we have the properties
$\xi_1=\xi_2^\ast$ and $\xi_3=\xi_3^\ast$. In view of the normal form
(\ref{dotxi2}) we introduce the abbreviations
\beq
r^2=\xi_1\xi_2\geq 0
\eeq
\beq
\label{abb}
g_j(r^2)=
    \sum_{n=1}^{[\hat{N}]}\;B_j(n\mu_1+n\mu_2+\mu_j)r^{2n}\;;
\;\;\;\hat{N}=\frac{\ep}{2|\rho_{11}|}\;;\;\;\;j=1,2,3
\eeq
\beq
h(r^2)=\sum_{n=1}^{[\tilde{N}]}\;B_3(n\mu_1+n\mu_2)r^{2n}\;;
\;\;\;\tilde{N}=\frac{\ep+|\la_3|}{2|\rho_{11}|}\;;
\eeq
\beq
a_k=2N_{k13}\;;\;\;\;k=1,2
\eeq
where $[\hat{N}]$ and $[\tilde{N}]$ denote the integer
parts of $\hat{N}$ and $\tilde{N}$, respectively. With this, in the
subspace $\Xi_1$ the normal form
(\ref{dotxi2}) has the following structure
\bdm
\dot{\xi}_{1} = [\la_1 + a_1\xi_3 + g_1(r^2)]\xi_1
\edm
\beq
\label{xi12}
\dot{\xi}_{2} = [\la_2 + a_2\xi_3 + g_2(r^2)]\xi_2
\eeq
\bdm
\dot{\xi}_3 = \la_3\xi_3 + h(r^2) + g_3(r^2)\xi_3
\edm
Obviously, the above system is decoupled from the remaining variables
$\xi_k$ with $k\geq 4$.

In the following we take the transformation (\ref{trans}) and with it the
normal form  (\ref{dotxi2})
up to cubic terms. With the aid of polar
coordinates $\xi_1=r\exp (i\fie)$ and by setting $\xi_3\equiv\xi$,
$\la_3\equiv\la$, $\rho_{11}\equiv\rho$ and $\om_{11}\equiv\om$ the
transformed dominant system is further reduced to effectively two dimensions
\beq
\label{dotr}
\dot{r}=r[\rho+\Re (a_1)\xi + br^2]\;\;;\;\;r\geq 0\;;
\eeq
\beq
\label{dotxi3}
\dot{\xi}=\la\xi + \ga_1 r^2 +  \ga_2 r^2\xi\;;
\eeq
\beq
\label{dotfie}
\dot{\fie}=\om +\beta r^2\;.
\eeq
As illustrated in fig. 2, we obtain a stable fixed point at $r=\xi=0$
and a further fixed point $F^\ast$ at $(r_c,\xi_c)$, which corresponds to a
Hopf limit cycle \cite{HassardKW} with angular frequency $\dot{\fie}$
given by (\ref{dotfie}).

Near the critical point $R=R_c$ we have $\xi_c=-\ga_1 r_c^2/\la$ which
renormalizes (\ref{dotr}) to $\dot{r}=r[\rho+b_L(R_c)r^2]$ with
the Landau--type parameter at $R=R_c$
\beq
\label{landauc}
b_L=b-\frac{\ga_1\Re (a_1)}{\la}\;.
\eeq
For higher order dominant systems there enter further renormalizing summands
into $b_L$.

{}From table 1 it is seen, that the Landau parameter $b_L$ of our normal form
method is almost independent of the choice of the interval parameter $\ep$,
as it should be.

\vspace{1cm}
Table 1\newline
Dependence of the Landau parameter $b_L$ on the choice of the interval
parameter $\ep$
at $R=5770$ with $E_1$ denoting the dimension of the corresponding dominant
system. The main results of the paper are for $E_1=3$.\\

\begin{tabular}{c|c|c|c}\hline
$\ep$       & $|\la_{q=0,\nu=2}|$  & $|\la_{q=0,\nu=1}| $  &
2$|\Re (\la_{q=1,\nu=1})|$ \\
$b_L$       & 39.163  &  39.168  & 39.174  \\
$E_1$       & 5  &  3  &  2 \\ \hline
\end{tabular}
\vspace{1cm}

As can be inferred from fig. 2, the basin of attraction of the
zero fixed point is limited by the curve $\dot{r}=0$.
If the starting point lies inside this basin of attraction, both the variables
of the dominant and the slaved system go to zero.

\vspace{1cm}

{\bf Here insert fig. 2}

\vspace{1cm}

Let us examine the slaved system at the non--zero fixed point
$(r,\xi)=(r_c,\xi_c)$. To this end we write down in the following the full
normal
form of our Galerkin system at $R=4320$ with the first two equations
referring to the dominant system (\ref{dotr}), (\ref{dotxi3})
\begin{eqnarray}
\label{domin1}
\dot{r}      & = & r(\rho      +  0.02942\xi    +  46.64r^2) \;\;\;;\;\;\;
r\geq0\\
\label{domin2}
\dot{\xi}    & = & \xi(\la  -  2.568r^2)       +  0.6561r^2  \\
\label{xi4}
\dot{\xi}_4  & = & \xi_4(\la_4 -  58.43r^2)       -  75.72\xi r^2  \\
\label{xi5}
\dot{\xi}_5  & = & \xi_5(\la_5 -  110.19r^2)      -  65.94\xi_4 r^2  \\
\dot{\xi}_6  & = & \xi_6(\la_6 +  80.37 r^2)                         \\
\dot{r}_1    & = & r_1(\rho_1  -  0.06202\xi    +  12.39r^2) \;\;\;;\;\;\;
r_1\geq0\\
\dot{\xi}_9  & = & \xi_9(\la_9 +  388.6 r^2)                         \\
\dot{r}_2    & = & r_2(\rho_2  -  0.08132\xi    +  83.36r^2) \;\;\;;\;\;\;
r_2\geq0\\
\dot{r}_3    & = & r_3(\rho_3  -  0.09851\xi    -  206.2r^2) \;\;\;;\;\;\;
r_3\geq0\\
\dot{\xi}_{14} & = & \xi_{14}(\la_{14} +  168.2r^2)                         \\
\dot{\xi}_{15} & = & \xi_{15}(\la_{15} -  155.95r^2)      +
2561\xi_4 r_1^2  \\
\dot{\xi}_{16} & = & \xi_{16}(\la_{16} -  61.77r^2)      -
86210\xi_4 r_2^2  \\
\dot{\xi}_{17} & = & \xi_{17}(\la_{17} +  14.79r^2)  +
15619\xi_5 r_3^2+63.38\xi_{14}r_1^2 \\
\dot{\xi}_{18} & = & \xi_{18}(\la_{18} +  0.6672r^2) + 19360\xi_{14}r_2^2 +
5671\xi_9r_3^2 \\
\dot{\xi}_{19} & = & \xi_{19}(\la_{19} -  1.430r^2)  +  1280\xi_{15}r_3^2  \\
\dot{\xi}_{20} & = & \xi_{20}(\la_{20} -  1.772r^2)  - 4.482\xi_{18}r_1^2 -
238.51\xi_{17}r_2^2  \\
\dot{\xi}_{21} & = & \xi_{21}(\la_{21} -  1.095 r^2) - 0.7456\xi_{19}r_1^2   \\
\label{xi22}
\dot{\xi}_{22} & = & \xi_{22}(\la_{22}  - 0.3739r^2) -172.1\xi_{19}r_3^2\;,
\end{eqnarray}
where the magnitudes $\rho=\Re (\la_1),\la=\la_3,\la_4,\ldots,\la_{22}$
of the linear problem are listed in Appendix D.
As in the case of the dominant system, polar coordinates are used for the
complex conjugate
amplitudes with $r_1$, $r_2$ and $r_3$ denoting the moduli of the modes
$(q=\pm 1,\nu=2)$, $(q=\pm 2,\nu=1)$ and $(q=\pm 3,\nu=1)$, respectively.
The equations for the corresponding polar angles are omitted.
The equations (\ref{xi4}) and (\ref{xi5}) have
nonzero fixed points $\xi_4^c$ and $\xi_5^c$, respectively,
 whereas each of the remaining equations
has fixed point zero.
Thus the overall nonzero fixed point at $R=4320$ is given by
\bdm
(r_c=0.008231,\xi_c=0.009170,\xi_4^c=-0.002647,
\edm
\beq
\xi_5^c=0.0003380, \xi_6^c=0,\ldots,\xi_{22}^c=0)\;.
\eeq
When the variables $r$ and $\xi$ are substituted by $r_c$
and $\xi_c$ in the slaved equations (\ref{xi4})--(\ref{xi22}),
 then it turns out that the separate fixed point of the slaved system
is stable. Therefore, at the critical point of the dominant subsystem there
is  no runaway of the slaved variables. We can thus define a meaningful
critical energy by mapping the coordinates of the nonzero fixed point back
into the original
phase space of the amplitudes $\eta$. With the aid of (\ref{ave}) we obtain
$E=0.26*10^{-3}$ at $R=4320$ which has to be compared with the unperturbed
energy
$E_0=8/15$ of the basic flow. The relative critical energy $E/E_0$ has the
order
of magnitude established in the literature, see e.g. \cite{Herbert}.

Regarding the fixed point in the original
phase space, we state the maximal amplitude for a given wave number $q$
\bdm
|\eta^{max}_{q=0}|=9.176*10^{-3}\;\;\;;\;\;\;|\eta^{max}_{q=1}|=5.726*10^{-3}
\edm
\beq
|\eta^{max}_{q=2}|=2.630*10^{-8}\;\;\;;\;\;\;|\eta^{max}_{q=3}|=
3.577*10^{-9}\;.
\eeq

As a final remark of this section, it is straightforward to extend the
given formalism to Reynolds numbers below $R=4320$ by choosing
a larger interval parameter $\ep$. This amounts to adding
further nonlinear terms to the normal form so as to preserve the same
structure in the larger Reynolds number interval.

\section{Comparison with Landau method}
Our main results are represented by fig. 3. We adopt the same normalization
\beq
\label{norm}
\Phi_{q=1,\nu=1}(z=0)=1
\eeq
as used in the Landau method \cite{HerbertL,Ven}.

\vspace{1cm}

{\bf Here insert fig. 3}

\vspace{1cm}

As is seen both the critical curve
of the Landau method  and of our normal form approach are consistent with the
neutral
curve of the Galerkin system, which is approximated by directly integrating
the
Galerkin system. The starting points were choosen close to the critical
amplitude
vector $\eta_s$, which is the image of the fixed point $F^\ast$ of the normal
form space.
This result also indicates that, in view of the Galerkin system choosen,
our function space of 22 Orr--Sommerfeld eigenfunctions is sufficiently large.
However the Galerkin space is to small to reproduce the Landau method results
near the critical point for $5760\leq R\leq R_c$.
We find $b_L(R_c)=39.17$, which is by $36\%$ larger than the corresponding
value,
 $28.88$, obtained by the Landau method \cite{Ven}; correspondingly
at the critical Reynolds number our  critical amplitude
is by $16\%$ smaller, see fig. 3 and fig. 4.
As should be remarked the critical curve of the Landau method \cite{Ven} is
established in the smaller interval $5300\leq R\leq R_c$.

For comparison with the results reported  elsewhere \cite{Ven} we give the
angular frequency  $\dot{\fie}$ of the nonlinear wave at the fixed
point $F^\ast$ for several Reynolds numbers. As table 2 shows,
the numerical values agree within $1\%$.

\vspace{1cm}
Table 2\newline
Comparison of the angular frequency of the nonlinear wave with the Landau
method.\\

\begin{tabular}{c|cccccc}\hline
R                               & 5200   & 5400   & 5500   & 5700   &
5770 \\ \hline
$\dot{\fie}$            & 0.2771 & 0.2744 & 0.2730 & 0.2704 & 0.2694     \\
$\dot{\fie}_{Landau}$    & 0.2743 & 0.2725 & 0.2717 & 0.2699 & 0.2693 \\\hline
\end{tabular}
\vspace{1cm}

In fig. 4 we show results for the parameter $b_L$ as a function of the
number $N$ of modes with $q=0$ considered.
The new boundary conditions, curve (2), and the usual ones, curve (1), give
practically the same value $b_L$.

\vspace{1cm}

{\bf Here insert fig. 4}

\vspace{1cm}

As to the definition domain of our normal form transformation we cannot safely
rely on the criterion of the Jacobian. In the given case the Jacobian is
larger $1$ well beyond the critical amplitude.

\section{ Conclusions}
A generalized normal form scheme for dissipative flows was presented
which allows to control the minimal magnitude of the normal form
denominators, and at the same time keeps a useful decoupling property
of the normal form. The decoupling property was rigorously proved and
constitutes
a new result. The feasibility of the method was examined for the well studied
example of plane
Poiseuille flow in the subcritical interval $4320\leq R\leq R_c$ of
the Reynolds number $R$. The critical energy
$E_c(R)$ or equivalently the critical amplitude $\eta_c(R)$, at which the
laminar flow becomes unstable, agrees well with the results obtained by other
methods \cite{Ven}, except close to the critical point where our critical
amplitude
is about $16\%$ smaller than the corresponding value of the Landau
method \cite{Ven}. The latter is established in the interval
$5300\leq R\leq R_c$,
whereas the present normal form gives reliable results down to $R=4320$,0 see
fig. 3.

The function space, in our case the Orr--Sommerfeld eigenfunctions,
is rigorously determined in principle for every parameter point $R$. For
comparison, both the Landau and the center manifold method rely on a critical
parameter
point $R_c$ of the linearized problem around which one expands in powers of
$\sqrt{R_c-R}$, for instance. The normal form method, on the other hand,
requires the existence
of a stable stationary solution only which provides a fixed point in phase
space.
Therefore the present method should be useful to study e. g. pipe flow or
plane Couette flow for which the linearized system lacks a critical point.

As a minor detail we studied boundary conditions which differ from the usual
ones for the modes with $q=0$.
Our new boundary conditions constitute a unified function space for all modes.
However it turns out that both boundary conditions give
practically the same results, see fig.4.

The quantitative validation of an existence domain is a notoriously difficult
problem in all three discussed methods.
We found our normal form results being consistent with results from the direct
integration of the Galerkin system. At the critical curve $r_c(R)$ we found
the Jacobian of the normal form transformation being always positive and well
away from zero. However, since our normal form transformation neglects higher
than cubic terms, the Jacobian criterion is not safe here.

The numerical efforts reside to about $90\%$ in the  normal form method,
whereas the remaining computing time lies mainly in establishing the
Orr--Sommerfeld
eigenfunctions and the Galerkin system.
For one parameter value $R$ the computing time was about
six minutes on a RISC 6000 work station.

The parameter interval $4320\leq R\leq R_c$ considered was determined by the
choice of the interval parameter $\ep$ together with the requirement that the
normal form (\ref{dotxi2})
keeps the same structure in the whole parameter interval. It is in principle
possible to go
deeper into the subcritical range by  choosing a larger $\ep$ which would
result in a higher dimensional dominant subsystem.
\section*{Appendix A: Recurrence relations}
We insert the normal form transformation (\ref{trans}) into the Galerkin
system (\ref{gals}) and consider the normalized dynamical system (\ref{dotxi})
as an ansatz. The left hand side of (\ref{gals}) then reads
\bdm
\dot{\eta}_j=\la_j\xi_j+\sum_{m\in M}[(m,\la)A_j(m)+B_j(m)]\xi^m
\edm
\beq
            +\sum_{m,m'\in M}\sum_lA_j(m')m_l'B_l(m-m'+\mu_l)\xi^m\;,
\eeq
where $\mu_k$ denotes the unit vector with
$(\mu_k)_i=\delta_{ik}$.
We use the convention that $A_j(m)$, $B_j(m)=0$ if
$m$ does not belong to the basic set $M$ defined in
(\ref{m}).
The right hand side of (\ref{gals}) becomes
\bdm
\la_j\xi_j+\la_j\sum_{m\in M}A_j(m)\xi^m +
\sum_{k,l}N_{j|kl}\left\{\xi^{\mu_k+\mu_l}
+\sum_{m\in M}[A_l(m-\mu_k)\right.
\edm
\beq
\left.
+A_k(m-\mu_l)]\xi^m
+\sum_{m,m'\in M}A_k(m-m')A_l(m')\xi^m\right\}\;.
\eeq
By equating the two sides, the linear terms cancel out.
The coefficients of the second order terms $\xi^m$
with $|m|=2$ read
\bdm
A_j(m)=\frac{1}{D_j(m)}\sum_{k,l}N_{j|kl}\delta(m,\mu_k+\mu_l)
\;\;\; {\rm if}\; m\notin {\cal R}_j(\ep)\;;
\edm
\beq
\label{rk2}
B_j(m) = -D_j(m)A_j(m) +
\sum_{k,l}N_{j|kl}\delta(m,\mu_k+\mu_l)\;\;\;\; {\rm if}\;\;
                                  m\in {\cal R}_j(\ep)\;,
\eeq
where $\delta(m,m')=1$ for $m=m'$ and
$\delta(m,m')=0$ for $m\neq m'$. Note that
$A_j(m)$ can be chosen arbitrarily if
$m\in {\cal R}_j(\ep)$.
In the case $|m|=3$ we get
\bdm
A_j(m)=\frac{1}{D_j(m)}\left[ 2\sum_{k,l}N_{j|kl}A_k(m-\mu_l)
\right.
\edm
\bdm
\left.
-\sum_{k=1}^n\;\sum_{m'\in {\cal R}_k(\ep)}B_k(m')[m_k-m'_k+1]
A_j(m-m'+\mu_k)\right]
\;\;\; {\rm if}\; m\notin {\cal R}_j(\ep)\;;
\edm
\bdm
B_j(m)=-D_j(m)A_j(m)+  2\sum_{k,l}N_{j|kl}A_k(m-\mu_l)
\edm
\beq
\label{rkn}
-\sum_{k=1}^n\;\sum_{m'\in {\cal R}_k(\ep)}\;B_k(m')[m_k-m'_k+1]
A_j(m-m'+\mu_k)
\;\;\; {\rm if}\; m\in {\cal R}_j(\ep)\;.
\eeq
Once more $A_j(m)$ can be chosen arbitrarily if $m\in {\cal R}_j(\ep)$.

For convenience we set the resonance coefficients $A_j(m)=0$.

\section*{Appendix B: Proof of decoupling theorem}

The forms (\ref{dynsys1}) and
(\ref{dynsys2}) of the normalized Galerkin system
imply that the dynamics of the subspace $\Xi_\si$
is independent of the higher subspaces $\Xi_{\si'}$
with $\si'>\si$ and depends only linearly on its
own variables if $\si\geq 2$.
We prove this by the method of contradiction
under the conditions (\ref{cond1})--(\ref{cond2}).
Let us consider the system $\dot{\Xi}_\si$, and assume that
i) in the normal form (\ref{dotxi2})  there is a
coefficient $B_{\si\tau}(m)\neq 0$ with
$m_{\si'\tau'}\neq 0$ and $\si'>\si$ or that
ii) there exists at least one coefficient
$B_{\si\tau}(m)\neq 0$ with
$\sum_{\tau'=1}^{E_\si}\;m_{\si\tau'}\geq 2$ for
$\si\geq2$.
In case i), because $|m|\geq 2$ always, and by (\ref{cond1})
the estimates are
\bdm
d_{\si\tau}(m)\geq |\Re (D_{\si\tau}(m))|\geq
|\rho_{\si'\tau'}+\rho_{11}-\rho_{\si\tau}|=
\edm
\beq
=|\rho_{\si'\tau'}-\rho_{\si\tau}|+|\rho_{11}|
\geq \De_{\si}+|\rho_{11}|>\ep\;,
\eeq
which tells that $m\notin {\cal R}_j(\ep)$ with $j\equiv (\si,\tau)$
and thus $B_{\si\tau}(m)=0$
in contradiction to the assumption i).\\
In the case ii) we get at first
\beq
d_{\si\tau}(m)\geq |\Re (D_{\si\tau}(m))|\geq
|\sum_{\si'=1}^\si \sum_{\tau'=1}^{E_{\si'}}m_{\si'\tau'}\rho_{\si'\tau'}
-\rho_{\si\tau}|\;.
\eeq
If  now
\beq
\label{casea}
|\sum_{\si'=1}^\si \sum_{\tau'=1}^{E_{\si'}}m_{\si'\tau'}\rho_{\si'\tau'}|
<|\rho_{\si\tau}|
\eeq
then by
diminishing the left hand side under the
assumption ii) we can write
\beq
2|\rho_{\si 1}|\leq |\rho_{\si\tau'}|+|\rho_{\si\tau''}|
\leq |\sum_{\si'=1}^\si \sum_{\tau'=1}^{E_{\si'}}m_{\si'\tau'}\rho_{\si'\tau'}|
<|\rho_{\si\tau}|\leq |\rho_{\si E_\si}|
\eeq
which implies $|\rho_{\si 1}|<W_\si$ and contradicts (\ref{cond2}).
In the case complementary to (\ref{casea}) we have
\beq
\label{caseb}
d_{\si\tau}(m)\geq
\sum_{\si'=1}^\si \sum_{\tau'=1}^{E_{\si'}}m_{\si'\tau'}|\rho_{\si'\tau'}|
-|\rho_{\si\tau}|\geq 2|\rho_{\si 1}|-|\rho_{\si E_\si}|\;.
\eeq
Making use of (\ref{cond2}) in the form $|\rho_{\si1}|>W_\si+\ep$ we
continue (\ref{caseb}) as
\beq
d_{\si\tau}(m)>|\rho_{\si 1}|+W_\si+\ep-|\rho_{\si E_\si}|=\ep
\eeq
which means that case ii) is not contained in the
resonant set ${\cal R}_j(\ep)$ with $j\equiv (\si,\tau)$.
This completes the proof.

\section*{Appendix C: Proof of slaving}

The dominant subspace $\Xi_1$ is independent
of the higher
subspaces $\Xi_\si$, $\si\geq2$. Since the
linear part of $\Xi_1$ is stable, the flow
$\dot{\Xi}_1$ possesses a finite basin of
attraction $BA$ by the theorem of Hartmann-Grobman
\cite{GuckenheimerH}. We claim that,
if the initial points $\Xi_1(t=0)\in BA$,
then the fixed point $\xi=0$ is stable
independently of the initial
values $\Xi_{\si}(t=0)$ with $\si\geq2$.
We demonstrate this in an exemplary way
for a three--dimensional system.
The general proof merely needs more efforts
in writing, but is otherwise
fully analogous.

Let us assume that the first subspaces
$\Xi_1(t),\ldots,\Xi_{\si-1}(t)$ are
stable, and that the next
subspace $\Xi_\si$ is three--dimensional
with one real variable $x$ and two
complex conjugate variables $r\exp(i\fie)$,
$r\exp(-i\fie)$ with $r\geq 0$.
By (\ref{dynsys2}) the corresponding
dynamical system reads
\beq
\label{xsys}
\dot{x}=[\la_{\si1}+\ga_1(t)]x
   +\ga_2(t)r+\theta_1(t)\;;
\eeq
\beq
\label{ysys}
\dot{r}=\ga_3(t)x+ [\rho_{\si2}+\ga_4(t)]r
+\theta_2(t)\;,
\eeq
where both $\rho_{\si2}=\Re (\la_{\si2})$ and $\la_{\si1}$
are negative, and
the real coefficients $\ga(t)$, $\theta(t)$ are determined
by the variables of the lower subspaces.
We need not know the phase $\fie=\fie(t)$. By
the inductive assumption, the four functions $\ga_1$ to $\ga_4$
and $\theta_{1/2}$ become arbitrarily
small for \( t \rightarrow \infty \).
Since the above system is linear with regular
time dependent coefficients,
the solutions $x(t)$ and $r(t)$ certainly
are bounded within any finite time interval.
With $\la=\max\{\la_{\si1},\rho_{\si2}\}<0$
the time derivative of the function
${\cal L}=x^2+r^2$ obeys the relation
\beq
\dot{{\cal L}}/2\leq \la [x^2+r^2]
+\ga_1x^2+\ga_4r^2 + [\ga_2+\ga_3]xr +\theta_1x+\theta_2r\;.
\eeq
Applying an orthogonal transformation
$(x,r)\rightarrow (y_1,y_2)$
we can get rid of the mixed term $xr$.
The transformed expression reads
\beq
\dot{{\cal L}}/2\leq \la' [y_1^2+y_2^2] +\tet_1y_1+\tet_2y_2\;,
\eeq
with $\la'=\la+\de(t)$ where $\de(t)$ and $\tet_{1/2}$ are arbitrarily
small magnitudes if $t$ is sufficiently large.
As a consequence, if we choose points with
\beq
|y_j|>\frac{2|\tet_j|}{|\la'|}\;\;;\;\;j=1,2\;,
\eeq
then $\dot{{\cal L}}<0$ and all initial
points end asymptotically
in the complementary, arbitrarily small
domain ${\cal D}_0$ with
$|y_j| \leq |2\tet_j/\la'|$.
As a final step we invoke the theorem of
Hartman-Grobman \cite{GuckenheimerH} which states
that our transformed
system (\ref{dynsys1}), (\ref{dynsys2}) is homeomorph to the
linearized problem in a finite neighbourhood of the origin,
provided all eigenvalues have a
negative real part.
If $t$ is sufficiently large, then ${\cal D}_0$ is a subset of
the validity domain of the Hartman-Grobman theorem. Thus,
if $t\rightarrow \infty$ then $(x,r)\rightarrow (0,0)$.

\section*{Appendix D: Table of eigenvalues}

 Table 2 \newline
 Real and imaginary part $\Re (\la)$, $\Im (\la)$ of the eigenvalues $\la_j$
 with
 $j=(q,\nu)$ for $R=4320$ and $R=5000$ at $\al=\al_c$. The ordering is given
 according to the real parts at $R=4320$.
 The different eigenvalues at given wave number q
 are labeled by $\nu=1,2,\ldots$. S and A denote symmetric and antisymmetric
 eigenfunctions, respectively.\newline
\begin{tabular}{l|r|r|c|c|c|c|c|}\hline
  No.  & $q$&$\nu$& S/A & $\Re (\la)$    & $\Im (\la)$ & $\Re (\la)$   &
$\Im (\la)$    \\
     &    &     &     & $R=4320 $     & $R=4320$    & $R=5000 $    &
$R=5000$ \\ \hline
1    &- 1 &1    &  S  & -0.0034297871 & 0.28323209  &-0.0015441660 &
0.27621304     \\
2    &  1 &1    &  S  & -0.0034297871 & -0.28323209 &-0.0015441660 &
-0.27621304  \\
3    &  0 &1    &  A  & -0.0046737834 &      0      &-0.0040381470 &      0
     \\
4    &  0 &2    &  A  & -0.013814702 &      0 &-0.011935903  &
     0        \\
5    &  0 &3    &  A  & -0.027523116 &      0 &-0.023779971  &      0      \\
6    &  0 &4    &  A  & -0.045800419 &      0 &-0.039571561  &
     0        \\
7    &- 1 &2    &  S  & -0.053931898 & 0.96621013 &-0.050157877  &
0.97004199  \\
8    &- 1 &2    &  S  & -0.053931898 & -0.96621013 &-0.050157877  &
-0.97004199    \\
9    &  0 &5    &  A  & -0.068646855 &      0 & -0.059310882 & 0    \\
10   &- 2 &1    &  A  & -0.075518479 & 1.9643698  &-0.070275192  &
1.9697702     \\
11   &  2 &1    &  A  & -0.075518479 & -1.9643698 &-0.070275192  &
-1.9697702   \\
12   &- 3 &1    &  S  & -0.090121159 & 2.9671465  &-0.084056136  &
2.9738632   \\
13   &  3 &1    &  S  & -0.090121159 & -2.9671465 &-0.084056136  &
-2.9738632   \\
14   &  0 &6    &  A  & -0.096062496 &    0    &-0.082997997  &    0    \\
15   &  0 &7    &  A  & -0.12804737 &    0    &-0.11063292  &    0    \\
16   &  0 &8    &  A  & -0.16460149 &    0    &-0.14221568  &    0    \\
17   &  0 &9    &  A  & -0.20572486 &    0    &-0.17774628  &   0    \\
18   &  0 &10   &  A  & -0.25141749 &    0   & -0.21722471 &    0    \\
19   &  0 &11   &  A  & -0.30167938 &    0   & -0.26065098 &    0    \\
20   &  0 &12   &  A  & -0.35651052 &    0   & -0.30802509 &    0    \\
21   &  0 &13   &  A  & -0.41591092 &    0   & -0.35934704 &    0    \\
22   &  0 &14   &  A  & -0.47988059 &    0   & -0.41461683 &    0    \\\hline
\end{tabular}

\vspace{0.8cm}
\large{\bf Acknowledgements}

This work was supported by the grant Ra 229/2 of
the Deutsche Forschungsgemeinschaft.
We are thankful to Dr.~Ludger Hannibal for helpful
discussions
and to Dr.~Charilaos
Kougias for giving computational advice in the beginning
of this work.

\newpage
\section*{Figure captions}
Fig. 1. Partitioning  of the eigenvalues into groups for the  decoupling
theorem at $R=4320$. The eigenvalues are indicated by circles.
The first stripe of width $W_1$ defines the dominant system; it
contains the two complex conjugate eigenvalues
which become critical at $R=R_c$ and one real eigenvalue.
All other stripes containing eigenvalues have zero width here.
The distances $\De_\si$ are all larger $\ep-|\rho_{11}|$, where $\ep$ is the
interval parameter of the normal form. Note that one easily could group the
eigenvalues in a different way when the distances become too small.
\\
\\
Fig. 2. Critical lines $\dot{r}=0$ and $\dot{\xi}=0$ in $\xi r$--space
as given by the transformed
dominant system (\ref{dotr}), (\ref{dotxi3}) for $R=4320$. The stable
fixed point is at the origin $(0,0)$, the unstable one is denoted by $F^\ast$.
The arrows indicate
the direction $(\dot{r},\dot{\xi})$ of the flow. For $r\geq 0.0427$, which is
out of scale,
the arrows along the curve $\dot{\xi}=0$ change sign.
\\
\\
Fig. 3. Critical amplitude $\eta_c$ (solid curve) by the normal form method
as a function
of the Reynolds number $R$,  and stability boundaries found by
 direct integration of the Galerkin system (dashed curves); the up and
down arrows indicate
the existence of growing modes  and decay of all modes, respectively.
The critical amplitudes of the normal form (1) and of the Landau method (2)
are compared in the insert.
\\
\\
Fig. 4. Dependence of the renormalized parameter $b_L$
on the number $N$ of the eigenfunctions with $q=0$ for $R=5770$.
Curve (1) and curve (2) refer to the usual and our  boundary conditions,
respectively.
In both cases the Orr-Sommerfeld basis functions are normalized to the same
energy.

\end{document}